\documentclass{PoS}

\newcommand{\eps}{\epsilon}

\newcommand{\et}{\tilde{e}}

\newcommand{\eh}{\hat{e}}
\newcommand{\Eh}{\hat{E}}
\newcommand{\Et}{\tilde{E}}

\newcommand{\nn}{\nonumber}

\newcommand{\bfc}{\begin{figure}\begin{center}}
\newcommand{\efc}{\end{center}\end{figure}}

\newcommand{\fig}[2]{\scalebox{#1}{\includegraphics{#2}}}

\newcommand{\Slash}[1]{\ooalign{\hfil/\hfil\crcr$#1$}}

\newcommand{\bra}[1]{\langle #1 |}
\newcommand{\ket}[1]{| #1 \rangle}

\newcommand{\Tr}[1]{{\rm Tr}\left[ #1 \right]}
\newcommand{\vb}{\biggl|}

\newcommand{\Oab}{\Omega^{\alpha}_{\,\,\,\,\beta}}
\newcommand{\psib}{\overline{\psi}}
\newcommand{\del}[2]{\frac{\partial #1}{\partial #2}}

\newcommand{\dZ}[1]{\frac{dz_{#1}}{z_{#1}^2}}
\newcommand{\lpartial}{\overleftarrow{\partial}}

\title{Contribution of the twist-3 fragmentation function to single
transverse-spin asymmetry in SIDIS}

\ShortTitle{Contribution of the twist-3 fragmentation function to SSA in SIDIS}

\author{\speaker{Koichi Kanazawa}\thanks{K.~K. is supported by JSPS
Research Fellowships for Young Scientists (No.24.6959).}\\
        Graduate School of Science and Technology, Niigata University,
        Ikarashi 2-8050, Niigata 950-2181, Japan\\
	Department of Physics, Barton Hall, Temple University,
        Philadelphia, PA 19122, USA\\
        E-mail: \email{kanazawa@nt.sc.niigata-u.ac.jp}}

\author{Yuji Koike\thanks{Y.~K. is supported in part by the Grant-in-
Aid for Scientific Research (No. 23540292) from the JSPS.}\\
        Department of Physics, Niigata University, Ikarashi 2-8050,
        Niigata 950-2181, Japan\\
        E-mail: \email{koike@nt.sc.niigata-u.ac.jp}}

\abstract{We study the contribution of the twist-3 fragmentation function
to the single transverse-spin asymmetry in SIDIS within the framework of
the collinear factorization. Using the Ward-Takahashi identity in QCD,
we establish the collinear twist-3 formalism in the Feynman gauge to
calculate its {\it non-pole} contribution to the asymmetry. 
}

\FullConference{XXI International Workshop on Deep-Inelastic Scattering
and Related Subject -DIS2013,\\
		22-26 April 2013\\
		Marseilles,France}
\begin{document}

\section{Introduction}

The single-transverse spin asymmetry (SSA) in high-energy QCD reactions
cannot be explained by the collinear parton model and its description
requires some extensions of the framework for QCD hard processes.
An extension is the twist-3 approach within the collinear factorization
framework where twist-3
parton distribution and/or fragmentation functions are responsible for
large SSAs in high-$P_T$ particle productions. 

So far, the collinear twist-3 approach has been mainly developed for the
pole contribution of the twist-3 distributions. On the other
hand, the study and knowledge of the twist-3 fragmentation has been scarce.
For the SSA in the $pp$
collision, a first calculation for the
so-called derivative contribution has been performed in
\cite{KangYuanZhou2010} while the complete formula has been derived in
the light-cone gauge in \cite{MetzPitonyak2013}. 
A numerical estimate of its impact in \cite{KangYuanZhou2010} shows this
effect could give a
significant contribution to the SSA, which suggests it could be a
possible origin of the
observed remarkably large SSA for $\eta$-meson at RHIC \cite{Star2012E}.
The importance of the twist-3 fragmentation effect has also been
pointed out in the context of the sign-mismatch
problem between the Sivers function extracted from semi-inclusive DIS
(SIDIS) and the
quark-gluon correlation function determined in the $pp$ process
\cite{KangQiuVogelsangYuan2011}.

In this report, we study the contribution of the twist-3 fragmentation
functions to the SSA in SIDIS.
A characteristic property of the twist-3 fragmentation 
 functions is that they could have complex phases to cause SSA in a
 nonperturbative way. Accordingly, for the calculation, we have to
 retain the {\it non-pole} part of
 the hard parts. So far, the color gauge-invariance of twist-3
 contributions in a covariant gauge is only shown for pole contributions
 \cite{EguchiKoikeTanaka2007,BeppuKoikeTanakaYoshida2010} while for the
 non-pole contribution the proof is absent in the literature.
Our main interest here is how the color gauge-invariance of the non-pole twist-3
contribution is realized in the Feynman gauge. 
For SIDIS, a first calculation has been made for the Collins azimuthal
asymmetry \cite{YuanZhou2009} although the cross-section formula for the
other azimuthal modulations has not been derived.
Establishing the collinear twist-3 formalism, we derive a complete
cross-section formula for the SSA in SIDIS.
A comparison of its prediction with the future EIC experiment would
shed new light on the origin of the SSA and multiparton correlations in
hadrons.

\section{Collinear twist-3 formalism for non-pole contribution}

We first introduce the F-type twist-3 fragmentation functions for a
spinless hadron $h$ as\,\cite{EguchiKoikeTanaka2006}
\begin{eqnarray}
%
%
\Delta_{F ij}^{\alpha} (z_1,z_2) &=& \frac{1}{N} \sum_X \int\frac{d\lambda}{2\pi} \frac{d\mu}{2\pi}
e^{-i\frac{\lambda}{z_1}} e^{-i\mu(\frac{1}{z_2}-\frac{1}{z_1})}
\bra{0} 
\psi_i(0) \ket{hX} 
\bra{hX} \psib_j(\lambda
w) 
gF^{\alpha\beta} (\mu w) w_{\beta} 
\ket{0} \nn\\
&=& \frac{M_N}{2z_2} (\gamma_5 \Slash{P}_h \gamma_{\lambda})_{ij}
\eps^{\lambda \alpha w P_h} \hat{E}_F (z_1,z_2) + \cdots,
\label{DeltaF} \\
%
%
\tilde{\Delta}_{F ij}^{\alpha} (z_1,z_2) &=& \frac{1}{N} \sum_X \int\frac{d\lambda}{2\pi} \frac{d\mu}{2\pi}
e^{-i\frac{\lambda}{z_1}} e^{-i\mu(\frac{1}{z_2}-\frac{1}{z_1})}
\bra{0} 
\psib_j (\lambda w) 
\psi_i(0) \ket{hX} 
 \bra{hX} gF^{\alpha\beta} (\mu w) w_{\beta} 
 \ket{0} \nn\\
&=& \frac{M_N}{2z_2} (\gamma_5 \Slash{P}_h \gamma_{\lambda})_{ij}
\eps^{\lambda \alpha w P_h} \tilde{E}_F (z_1,z_2) + \cdots, \label{DeltaFt} 
\end{eqnarray}
where $\psi_i$ is the quark field with spinor index $i$ while
$F^{\alpha\beta} \equiv F^{a\alpha\beta} T^a$ is the gluon's field
strength with $T^a$ being the color matrices.
$M_N$ is the nucleon mass, $w^\mu$ is a light-like vector satisfying
$P_h\cdot w=1$, $N=3$ is the number of colors and $\eps^{\lambda\alpha w
P_h} \equiv \eps^{\lambda\alpha \rho\sigma} w_{\rho} P_{h\sigma}$ with the
Levi-Civita tensor being $\eps_{0123} \equiv +1$. We have suppressed the
gauge-link operators between the fields for simplicity.
$\{ \Eh_F, \Et_F \}$ is a complete set of the twist-3
quark-gluon correlation functions for the spinless hadron.
These are chiral-odd functions and enter the single-spin dependent
cross-section formula with the quark transversity distribution.

As is well-known, the F-type function
 $\Eh_F(z_1,z_2)$ is related with the D-type
function $\Eh_D(z_1,z_2)$, which is defined with the covariant
 derivative $D^\alpha$
 instead of $gF^{\alpha\beta}w_\beta$,
 as\,\cite{EguchiKoikeTanaka2006}
\begin{eqnarray}
\Eh_D (z_1,z_2) = P \left( \frac{1}{1/z_1-1/z_2} \right)
\Eh_F (z_1,z_2) + 
\delta\left(\frac{1}{z_1}-\frac{1}{z_2} \right) \et (z_2),\label{relation}
\end{eqnarray}
where $\et (z)$ is given by\footnote{Note the F-type function
vanishes at the soft-gluon pole point:
$\Eh_F(z,z)=0$\,\cite{MeissnerMetz2009,GambergMukherjeeMulders2011}.}
\begin{eqnarray}
 \Delta_{\partial ij}^\alpha (z) &=& \frac{1}{N} \sum_X \int
  \frac{d\lambda}{2\pi} e^{ -i\frac{\lambda}{z} } \bra{0} [\infty w,
  0] \psi_i(0)
   \ket{hX} \bra{hX} \psib_j (\lambda w)
  [\lambda w,\infty w]
  \ket{0} \lpartial^\alpha \nn\\
 &=& \frac{M_N}{2z} (\gamma_5 \Slash{P_h}\gamma_\lambda)_{ij}
  \epsilon^{\lambda\alpha wP_h} \et (z) + \cdots. \label{Deltapartial}
\end{eqnarray}
Here we restore the gauge-link operators to emphasize the derivative
$\lpartial^\alpha$ hits both $\psib(\lambda w)$ and $[\lambda w, \infty w]$.
For the calculation, we also need the
2-parton correlator  \cite{Ji1994}
\begin{eqnarray}
 \Delta_{ij}(z) &=& \frac{1}{N}\sum_X \int \frac{d\lambda}{2\pi} \bra{0}
  \psi_i(0) \ket{hX}\bra{hX} \psib_j(\lambda w)\ket{0} \nn\\
 &=& \frac{M_N}{2z}
  (\sigma_{\lambda\sigma}i\gamma_5)_{ij} \epsilon^{\lambda\sigma
wP_h} \eh_{\overline{1}}(z)+\cdots.
\end{eqnarray}
This twist-3 function $\eh_{\overline{1}}$ is related to the imaginary part of the
D-type function via the QCD equation of motion \cite{MetzPitonyak2013}
and thus is expressed in terms of $\Eh_F$ and $\et$ through
Eq.\,(\ref{relation}).

\begin{figure}[t]
 \begin{center}
  \fig{0.4}{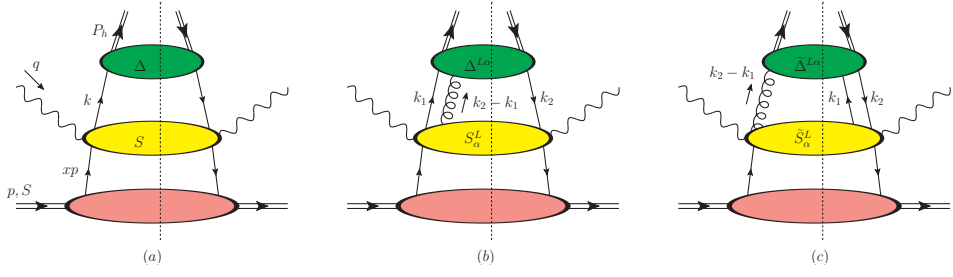}
  \caption{Generic diagrams giving rise to the twist-3 fragmentation
  function contribution to the asymmetry. The top blobs are the
  fragmentaion matrix elements, the middle blobs are the hard parts, and
  the bottom blobs represent the quark transversity. The mirror diagrams
  of $(b)$ and $(c)$ also contribute. \label{generic}}
 \end{center}
\end{figure}

The single-spin dependent cross-section in SIDIS may be written in terms of the
leptonic and hadronic tensors, $L^{\mu\nu}$ and $W_{\mu\nu}$,
as $\Delta\sigma \sim L^{\mu\nu} W_{\mu\nu}$ \footnote{For the kinematics, see \cite{KoikeTanakaYoshida2011}.}.
For the analysis of the twist-3 fragmentation effect, we factorize the
quark transversity distribution $h(x)$ from the hadronic tensor,
\begin{eqnarray}
W_{\mu\nu} = \int \frac{dx}{x} h (x) w_{\mu\nu}.
\end{eqnarray}
The generic diagrams which contain twist-3 fragmentation
effects are shown in Fig.\,\ref{generic}. 
We first consider the diagrams $(a)$, $(b)$ and the mirror diagram of
$(b)$. A standard and systematic way to extract twist-3 effects is the
collinear expansion of the hard parts. Following the procedure in
\cite{EguchiKoikeTanaka2007}, it is easy to see the twist-3 contribution is
expressed in terms of the gauge-invariant F-type matrix element
(\ref{DeltaF}) and several gauge-dependent ones. 
%
As is well-established, for pole contributions, the latter 
 gauge-dependent terms vanish owing to the Ward identites for the hard
 parts which hold due to the on-shell conditions associated with the
 poles of the internal propagators
 \cite{EguchiKoikeTanaka2007,BeppuKoikeTanakaYoshida2010}.
Thus the color gauge-invariance is ensured for the pole contribution at
 twist-3.
For the non-pole contribution, however, such special on-shell conditions
 are lacking, so that it is necessary to develop the collinear twist-3
 formalism without relying on those conditions. 

\begin{figure}
 \begin{center}
  \fig{0.6}{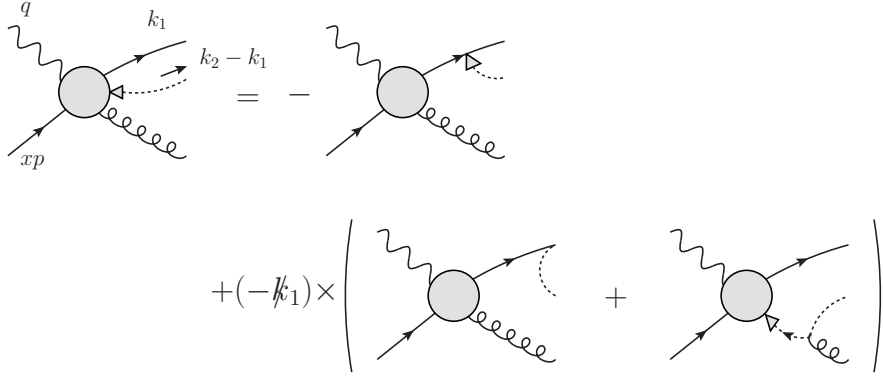}
 \end{center}
 \caption{Ward-Takahashi identity for a coupling of the
 scalar-polarized gluon $(k_1-k_2)^\sigma$ onto the amplitude in the
 left-hand-side of the
 final-state cut in $S^{La}_\sigma (k_1,k_2)$. 
The circle
 represents the sum of the $s$-channel and $t$-channel diagrams for the
 partonic subprocess $\gamma^* q\to qg$. 
 \label{wti}}
\end{figure}

In order to show the desired factorization property for the non-pole contribution,
what we have to do is to prove all those
gauge-dependent terms are combined into matrix elements for the
gauge-invariant operators.
To show this is the case, we note the hard
part $S^L_\sigma$ with an off-shell quark leg obeys a Ward-Takahashi
identity, as shown in Fig.\,\ref{wti},
\begin{eqnarray}
(k_2-k_1)^\sigma S^{La}_\sigma (k_1,k_2) = T^a S (k_2). \label{hwti}
\end{eqnarray}
Here we discarded the ghost-like term, the last diagram in the
right-hand-side in
Fig.\,\ref{wti}. This can be done in the twist-3 accuracy because the
hard part for the ghost-like term
and its first derivatives with respect to $k_{1,2}$ vanish in the
collinear limit. 
With the identity (\ref{hwti}), we find the light-cone
matrix elements
produced in the course of the collinear expansions are eventually
reorganized into the color gauge-invariant ones as\footnote{We have
suppressed the Lorentz indices $\mu$ and $\nu$ for simplicity.}
\begin{eqnarray}
w^{(a)} + w^{(b)} + w^{(b)*} &=& \int \dZ{} \Tr{\Delta(z)S(z)} + \Oab \int \dZ{} {\rm Im} \Tr{
  \Delta_\partial^\beta (z) \del{S(k)}{k^\alpha} \vb_{k\to\frac{P_h}{z}}
  } \nn\\
&& - 2 \Oab \int \dZ{1} \dZ{2} P \left( \frac{1}{1/z_2 - 1/z_1} \right)
{\rm Im} \Tr{ \Delta_F^\beta (z_1,z_2) S_\alpha^L (z_1,z_2)
  }, \label{wabc}
\end{eqnarray}
with the projection operator $\Omega^{\alpha\beta} = g^{\alpha\beta} -
P_h^\alpha w^\beta$. Note in this analysis we have recovered up to the ${\cal
O} (gA)$ terms including the expansion of the gauge-link operators in
Eq.\,(\ref{Deltapartial}).

For completeness, we have to take into account the contributions from
the diagram $(c)$ in Fig.\,\ref{generic} and its mirror diagram.
In this case, the hard part satisfies a simple Ward identity, 
\begin{eqnarray}
 (k_2-k_1)^\sigma \tilde{S}^L_\sigma (k_1,k_2) = 0.
\end{eqnarray}
Since this is the same with the case in pole contribution, it is
straightforward to show the factorization property and the color
gauge-invariance. The resultant
factorization formula is expressed in terms of only the gauge-invariant
F-type matrix element (\ref{DeltaFt}) as
\begin{eqnarray}
 w^{(c)} + w^{(c)*} = - 2 \Oab \int \dZ{1} \dZ{2} P \left(
					       \frac{1}{1/z_2-1/z_1}
					      \right) {\rm Im}
 \Tr{\tilde{\Delta}_F^\beta (z_1,z_2) \tilde{S}_\alpha^L (z_1,z_2)}. \label{wde}
\end{eqnarray}
Equations \,(\ref{wabc}) and (\ref{wde}) show all contributions
arising from the twist-3 fragmentation functions have definite
factorization property with manifest color gauge-invariance.
The complete partonic cross-section formula can be found in
\cite{KanazawaKoike2013F}.

\section{Summary}

We have discussed the
contribution of the twist-3 fragmentation function
to the SSA in SIDIS in the framework of the collinear factorization. We
have established the collinear
twist-3 formalism in the Feynman gauge to derive the gauge-invariant
factorized cross-section formula.
There, the relations among the hard
parts by the Ward-Takahashi identity play a crucial role in
reorganizing the twist-3
light-cone matrix elements into the color gauge-invariant ones.
The single-spin dependent cross-section formula is expressed in terms of
the gauge-invariant functions, $\Eh_F,\Et_F$, and $\et$.
Future EIC experiments would
provide us with a unique opportunity to determine these functions and
clarify the origin of large SSAs.


\end{document}